\documentclass[12pt]{article}

\usepackage[a4paper,top=2.5cm,left=1.75cm,right=1.75cm,bottom=3.5cm]{geometry}
\usepackage[utf8]{inputenc}
\usepackage[T1]{fontenc}
\usepackage{authblk}
\usepackage{amsmath,amssymb,amsfonts,amsthm}
\usepackage{lineno}
\usepackage[table,svgnames,dvipsnames]{xcolor}
\usepackage{graphicx}
\usepackage{tabularx}
\usepackage[colorlinks=true, linkcolor=black, urlcolor=blue, citecolor=blue]{hyperref}
\usepackage{booktabs}
\usepackage[labelfont=bf,margin={1.75cm,1.75cm},font=small]{caption}
\usepackage{rotating}
\usepackage{multirow}
\usepackage{titlesec}

\usepackage[maxnames=3,minnames=3,giveninits=true,sorting=none,style=nature]{biblatex}
\title{Extending the RANGE of Graph Neural Networks:\\Relaying Attention Nodes for Global Encoding}

\author[1]{Alessandro Caruso\textsuperscript{\textdagger}}
\author[1,2]{Jacopo Venturin\textsuperscript{\textdagger}}
\author[1]{Lorenzo Giambagli\textsuperscript{\S}}
\author[1]{Edoardo Rolando\textsuperscript{\S}}
\author[3,2,1,5]{Frank No{\'e}\textsuperscript{*}}
\author[1,4,5]{Cecilia Clementi\textsuperscript{*}}

\affil[1]{Department of Physics, Freie Universit\"at Berlin, \emph{Arnimallee 12}, 14195, Berlin, Germany}
\affil[2]{Department of Mathematics and Computer Science, Freie Universit\"at Berlin, \emph{Arnimallee 12}, 14195, Berlin, Germany}
\affil[3]{Microsoft Research AI for Science, \emph{Karl-Liebknecht Str. 32}, 10178, Berlin, Germany}
\affil[4]{Center for Theoretical Biological Physics, Rice University, Bioscience Research Collaborative, \emph{6500 Main Street}, Houston, 77005, TX, USA}
\affil[5]{Department of Chemistry, Rice University, \emph{6100 Main Street}, Houston, 77030, TX, USA}

\date{}

\begin{document}

\maketitle

\begin{center}
\textsuperscript{*}Corresponding authors. E-mails: \href{mailto:frank.noe@fu-berlin.de}{frank.noe@fu-berlin.de}, \href{mailto:cecilia.clementi@fu-berlin.de}{cecilia.clementi@fu-berlin.de}\\
\textsuperscript{\textdagger}These authors contributed equally.\\
\textsuperscript{\S}These authors contributed equally.\\
\end{center}

\abstract{Graph Neural Networks (GNNs) are routinely used in molecular physics, social sciences, and economics to model many-body interactions in graph-like systems. However, GNNs are inherently local and can suffer from information flow bottlenecks. This is particularly problematic when modeling large molecular systems, where dispersion forces and local electric field variations drive collective structural changes. Existing solutions face challenges related to computational cost and scalability. We introduce RANGE, a model-agnostic framework that employs an attention-based aggregation-broadcast mechanism that significantly reduces oversquashing effects, and achieves remarkable accuracy in capturing long-range interactions at a negligible computational cost. Notably, RANGE is the first virtual-node message-passing implementation to integrate attention with positional encodings and regularization to dynamically expand virtual representations. This work lays the foundation for next-generation of machine-learned force fields, offering accurate and efficient modeling of long-range interactions for simulating large molecular systems.}

\section*{Introduction}
In the last decade, Message Passing Neural Networks (MPNNs) and, more generally, Graph Neural Networks (GNNs) have been established as a powerful and flexible approach to learning from graph-structured data~\supercite{scarselli2008graph,kipf2016semi,battaglia2018relational}.
In GNNs, the graph nodes take the role of artificial neurons, and local many-body information is aggregated in each message-passing step by updating the node weights with messages received from direct neighbor nodes. By repeating such message-passing steps multiple times, the field of view of each node expands to higher-order neighbors.

In molecular science, GNNs have been found particularly useful in the development of Machine-Learned Force-Fields (MLFFs), where the nodes correspond to particles with a physical location in three-dimensional space - either corresponding to atoms in an atomistic force-field~\supercite{schutt2017schnet,schutt2018schnet,schutt2021equivariant,unke2021spookynet,batatia2022mace,batzner2022,frank2022so3krates}, or beads in a coarse-grained (CG) force-field~\supercite{husic2020coarse,charron2023navigating,durumeric2023machine,kramer2023statistically}. These MLFFs are trained using energies or forces of molecules and configurations coming from a trusted ground-truth, such as quantum chemistry calculations or classical all-atom simulations.
MLFFs have evolved in the past years, reflecting new trends and the fast development of network architectures in machine learning. 
Examples include the incorporation of physical symmetries and equivariances~\supercite{schutt2017schnet,batatia2022mace}, attention mechanisms~\supercite{frank2022so3krates,tholke2022torchmd,frank2024euclidean}, and the integration of physics-based functional forms~\supercite{unke2021spookynet,kabylda2024molecular}.

The main limitation of GNN-based MLFFs is that they are inherently local. The neighborhood of each particle node is usually defined to be all the other particles within a cutoff radius. In each message-passing step, information is exchanged within this radius. The field of view of each graph node is thus limited by the cutoff radius multiplied by the number of message-passing steps. While most MLFFs use cutoff radii of a few \AA ngstr\"oms to limit the computational cost of the message-passing operations, long-ranged electrostatic interactions can span several tens of \AA ngstr\"oms, in particular at interfaces such as biomembranes or in low-dielectric solvents~\supercite{rossi2015stability,stohr2019quantum}.

The brute-force approach of extending the number of message-passing steps leads to highly correlated node representations, averaging out the information, that, as it travels across the network, is further deteriorated by the presence of topological bottlenecks~\supercite{alon2020bottleneck}. These two well-known limitations of GNNs with many message-passing steps and large cutoffs, respectively known as oversmoothing and oversquashing, significantly impair long-range message-passing. Moreover, extending the cutoff radius so that the field-of-view covers the entire system size, requires the evaluation of $O(N^2)$ interactions for a system of $N$ particles, leading to computational costs at inference, and to memory costs during training, which become prohibitive when scaling to large particle numbers.

Several solutions have been proposed to address long-range interactions in MLFFs. In classical molecular dynamics (MD), long-range interactions in periodic systems are typically treated using Ewald summation~\supercite{toukmaji1996ewald}. Inspired by that, Ewald message-passing combines a direct-interaction GNN between particles in real space with a network in the Fourier representation of the periodic particle density~\supercite{kosmala2023ewald,geisler2024spatio,loche2024fast,wang2024neural}. Despite the use of Fast Fourier Transforms (FFTs)~\supercite{loche2024fast,wang2024neural}, these methods are quite computationally expensive.
Another way to enable a global field of view while avoiding oversquashing is to employ global self-attention for each node. Inspired by Large Language Models (LLMs), where its effectiveness is well established~\supercite{vaswani2017attention}, this approach updates node representations by aggregating information from the entire graph via a weighted average of the constituent nodes, with the normalized weights calculated from each node-pair~\supercite{velivckovic2017graph,brody2021attentive}. The main drawback of global attention is its high computational and memory cost, which scale as $O(N^2)$.
By introducing a series of approximations, memory requirements can be significantly reduced\supercite{dao2022flashattention}, enabling linear time scaling~\supercite{choromanski2020rethinking,dao2023flashattention,shah2024flashattention}. In this direction, notable progress has also been achieved in the atomistic domain~\supercite{unke2021spookynet,kong2023goat,frank2024euclidean}.
Lastly, the addition of virtual graph elements offers a straightforward method to extend message-passing across the entire graph. Although this concept was first introduced in molecular physics almost a decade ago~\supercite{gilmer2017neural}, its adoption has been relatively limited~\supercite{li2024neural}, despite its demonstrated success in other fields~\supercite{li2017learning,pham2017graph,ishiguro2019graph,ye2019bp,sestak2024vn}. Virtual nodes that aggregate and broadcast information to the entire structure are particularly appealing, as they are characterized by linear time complexity and it has been theorized that they can approximate a self-attention mechanism with some assumptions on the structure of the virtual representation~\supercite{cai2023connection}; however, previous implementations were architecture-dependent, using the same message-passing algorithm as the underlying model, and represented the entirety of the system with a single fixed-size vector, limiting the flow of information in the case of arbitrarily large structures~\supercite{alon2020bottleneck}.

In this work, we present RANGE (Relaying Attention Nodes for Global Encoding): an extension to GNN architectures that can be flexibly combined with a large variety of base frameworks, achieving long-range many-body message-passing for graphs of arbitrary topology. In contrast to existing approaches, RANGE introduces multiple virtual representations with positional encodings that relay information via self-attention, strongly reducing oversmoothing and oversquashing and scaling linearly with system size.

\begin{figure}[h!]
    \centering
    \includegraphics[width=0.9\textwidth]{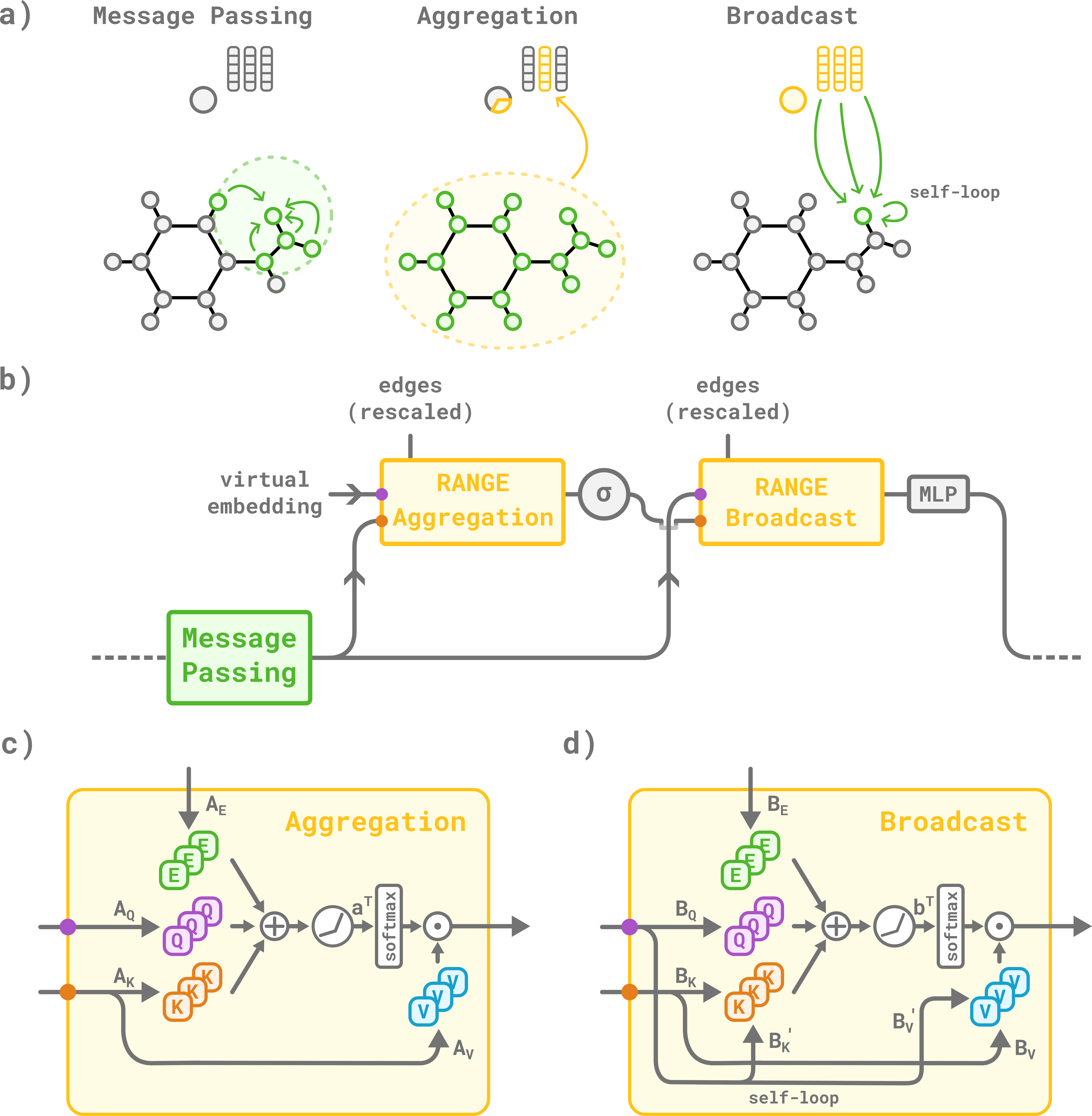}
    \caption{\textbf{Overview of RANGE.} In a) and b), after the message-passing step, the updated node representation and the initialized virtual embeddings are fed into the RANGE aggregation block. After an element-wise non-linearity, the coarse-grained representation is propagated back via the RANGE broadcast block. The mixing between different heads is done by a multilayer perceptron. In c) and d), aggregation and broadcast blocks project senders and receivers onto key and query space respectively. A positional encoding projected onto the edge space is included in the calculation of the attention weights. During the broadcast phase d), a memory effect, modeled by self-loops, is introduced for balancing local and global information content inside each graph node.}
    \label{fig:architecture}
\end{figure}
\section*{Results}

\subsection*{Overview of RANGE}
Building on the standard MPNN paradigm, RANGE introduces a set of virtual nodes as global representations of the underlying graph, to which we refer as master nodes (Fig.~\ref{fig:architecture}). After a standard message-passing step, during the \textit{aggregation} phase, node embeddings are gathered into coarse-grained representations via multi-head self-attention, producing independent representations of aggregated information. This information is distributed back to the graph nodes during the \textit{broadcast} phase; the nodes of the base graph can weigh the relative impact of individual master nodes, while preserving relevant information collected during the message-passing step thanks to the presence of self-loops.
Since the master nodes have direct edges to every node of the graph, they capture long-range interactions in a single step, overcoming limitations of strictly local, pairwise, receptive fields, and simultaneously avoid the oversmoothing that would come with repeating many message-passing steps and the oversquashing that stems from transmitting information through a single finite-dimensional channel, effectively compressing the flow of information. The presence of master nodes dramatically changes the topology of the graph towards a \textit{small world} structure, in which information can travel long distances with only a few steps~\supercite{watts1998collective}. Refer to the Methods section and Supplementary Note~\ref{sinote:range} for a detailed description of RANGE.

\subsection*{Accuracy and Computational Cost}
RANGE is an architectural extension that can, in principle, be applied on top of any message-passing framework.
Among the state-of-the-art MPNNs, SchNet~\supercite{schutt2017schnet,schutt2018schnet} and PaiNN~\supercite{schutt2021equivariant} have become popular frameworks for modeling molecular systems~\supercite{husic2020coarse,jorgensen2022equivariant,charron2023navigating,schreiner2024implicit}. While the former utilizes invariant node representations, the latter also employs equivariant embeddings, leading to higher accuracy in the prediction of both invariant and equivariant properties with a higher computational cost~\supercite{frank2023peptides}. Here, we use both SchNet and PaiNN as baseline models to perform extensive analyses and demonstrate the performance of RANGE in terms of accuracy and efficiency.
\begin{figure}[h!]
    \centering
    \includegraphics[width=0.9\textwidth]{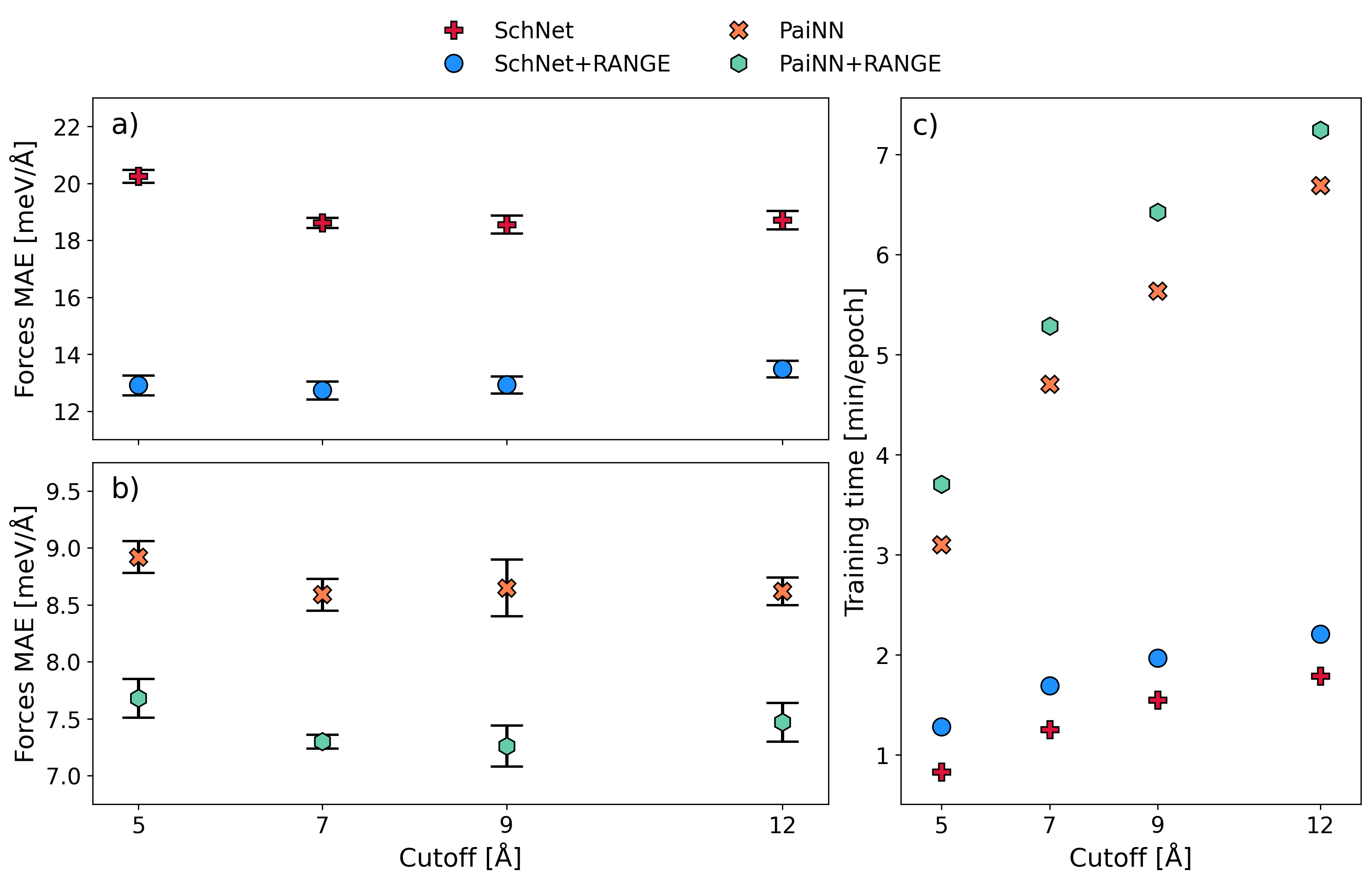}
    \caption{\textbf{Accuracy and training time dependence on message-passing cutoff.}
    The MAE on the predicted forces for the AQM dataset is reported for a) SchNet and b) PaiNN, and the same models with the RANGE extension, as a function of the message-passing cutoff. In c), the training time per epoch is reported for the same models. All the presented values are averaged on 4 models independently trained with different dataset seeds.}
    \label{fig:accuracy_timing}
\end{figure}
We apply RANGE to train atomistic MLFFs on two different datasets to cover different molecular environments and system sizes: QM7-X~\supercite{hoja2021qm7}, comprised of relatively small structures with up to 23 atoms, and Aquamarine (AQM)~\supercite{medrano2024dataset}, representing more challenging and interesting structures, ranging from 30 to 92 atoms. As we further explain in Supplementary Note~\ref{sinote:dataset}, the reference data explicitly include the accurate quantum treatment of long-range effects via many-body dispersion~\supercite{tkatchenko2012accurate,ambrosetti2014long}.

We first compare the mean absolute error (MAE) of energy and forces with respect to the training time per epoch for the AQM dataset using both SchNet and PaiNN, and their RANGE counterparts, using 3 interaction layers and different cutoff values (Fig.~\ref{fig:accuracy_timing}, a and b; Supplementary Note~\ref{sinote:model_training}).
We find that RANGE consistently outperforms the SchNet and PaiNN baseline models at any chosen cutoff. For both baseline models, increasing the cutoff only slightly increases their performance; around 9-12\,\AA\, the error saturates or even slightly increases, indicating the presence of information bottlenecks, i.e. oversquashing. On the other hand, even the RANGE models with the shortest cutoff outperform the baseline models with longest reach. This leads to a significant saving in computational cost: at any given cutoff, the training time per epoch of RANGE increases only slightly over the baseline model (Fig.~\ref{fig:accuracy_timing}, c).
The energy prediction and all the numerical values are reported in Supplementary Fig.~\ref{sifig:accuracy_timing} and Supplementary Table~\ref{sitab:aqm_painn_schnet}, respectively.

\begin{figure}[h!]
    \centering
    \includegraphics[width=0.9\textwidth]{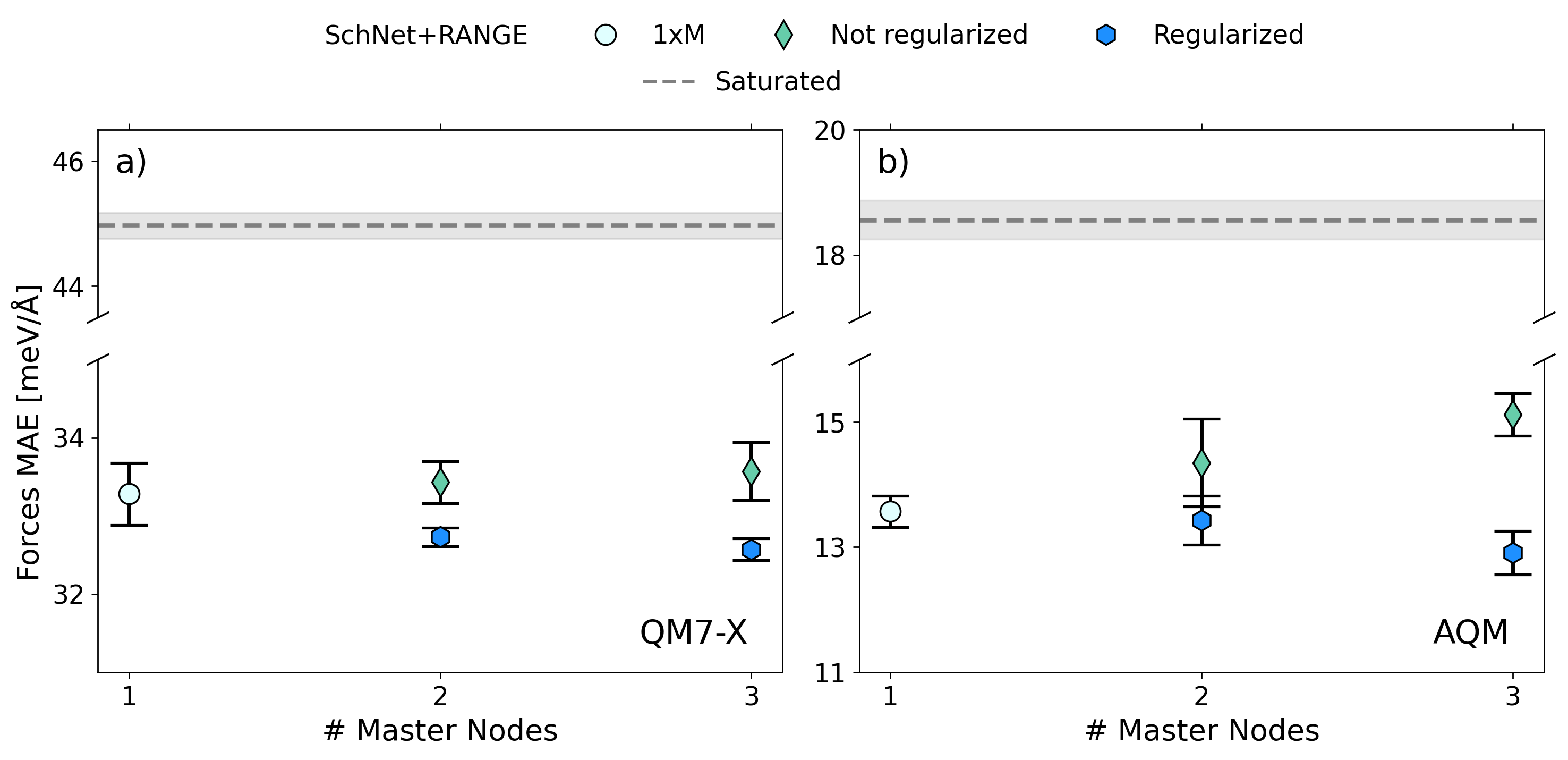}
    \caption{\textbf{MAE of the regularized and non-regularized RANGE model.} Force MAE of the regularized and non-regularized RANGE models with different number of master nodes are reported for the a) QM7-X and b) AQM datasets. The gray line represents the lowest  MAE achieved by the baseline model upon increasing the message-passing cutoff. All the reported values are averaged on 4 models independently trained with different dataset seeds.}
    \label{fig:regularized}
\end{figure}
While it was recently suggested that adding global aggregations could only lead to better performances~\supercite{li2024neural}, we observe that, if these are left unconstrained, the attention weights of multiple master nodes can become degenerate (Supplementary Fig.~\ref{sifig:regularization_weights}), leading to a degradation of accuracy with an increasing number of master nodes (Fig.~\ref{fig:regularized}; Supplementary Table~\ref{sitab:qm7x_aqm_dha_general}). To address this issue, we introduce a regularization procedure to dynamically allocate the number of master nodes as a function of the system size, effectively acting as an expandable space for storing global information (Supplementary Note~\ref{sinote:range}; Supplementary Fig.~\ref{sifig:not_regularized_accuracy_timing}). We stress that, for both datasets in Fig.~\ref{fig:regularized}, all RANGE models lie below the smallest possible MAE that is achievable by naively increasing the message-passing cutoff, even in QM7-X, where the large majority of compounds is fully included within the largest cutoff value tested (7\,\AA).

\begingroup
\renewcommand{\arraystretch}{1.125}
\begin{table}[h!]
    \centering
    \caption{\textbf{Comparison between Ewald MP and RANGE.} MAE of energy and forces, and relative training time per epoch of the AQM dataset are reported for Ewald MP, RANGE, and the respective SchNet and PaiNN baseline models. All the reported values are averaged on 4 models independently trained with different dataset seeds.}
    \label{tab:comparison-ewald-range}
    \begin{tabular}{ll*{3}{c}}
        \toprule
         & \multirow{2}{*}{Model} & MAE energy & MAE forces & Rel. training time\\
         &       & [meV] & [meV/\AA] & [a.u.]\\
        \midrule
         \parbox[t]{2mm}{\multirow{3}{*}{\rotatebox[origin=c]{90}{SchNet}}} & Baseline & $46.6 \pm 1.1$ & $20.3 \pm 0.2$ & - \\
         & Ewald MP & $45.6 \pm 0.6$ & $19.3 \pm 0.1$ & $3.851 \pm 0.017$ \\
         & \cellcolor{Gainsboro!60}RANGE & \cellcolor{Gainsboro!60}$\mathbf{27.8} \pm 1.4$ & \cellcolor{Gainsboro!60}$\mathbf{12.9} \pm 0.4$ & \cellcolor{Gainsboro!60}$\mathbf{1.540} \pm 0.008$ \\
        \midrule
         \parbox[t]{2mm}{\multirow{3}{*}{\rotatebox[origin=c]{90}{PaiNN}}} & Baseline & $24.5 \pm 0.7$ & $8.9 \pm 0.1$ & - \\
         & Ewald MP & $23.3 \pm 1.1$ & $8.8 \pm 0.2$ & $2.290 \pm 0.010$ \\
         & \cellcolor{Gainsboro!60}RANGE & \cellcolor{Gainsboro!60}$\mathbf{19.5} \pm 0.5$ & \cellcolor{Gainsboro!60}$\mathbf{7.7} \pm 0.2$ & \cellcolor{Gainsboro!60}$\mathbf{1.197} \pm 0.004$ \\
        \bottomrule
    \end{tabular}
\end{table}
\endgroup

As a notable example among Ewald-based methods, Ewald MP~\supercite{kosmala2023ewald} projects the node embeddings onto the reciprocal space via Fourier expansion and applies a learned frequency filter to specifically select long-range interactions; after transforming the embeddings back to the real space, the additional contribution is added to the prediction of the baseline model. Since both Ewald MP and RANGE can be applied to virtually any MPNN out-of-the-box, we compare their performances for SchNet and PaiNN with a 5\,\AA\, cutoff on the AQM dataset (Table~\ref{tab:comparison-ewald-range}). Not only RANGE achieves a drastically lower MAE with respect to both the baseline models and Ewald MP, but its application also comes at a significantly lower computational cost with respect to Ewald MP due to the inexpensive prefactor and better time scaling.

\subsection*{Molecular Dynamics Simulations with RANGE}
An important requirement for atomistic force-fields is the continuity of energies and forces with respect to the positions of the input coordinates. This property is well-known and is often obtained in standard MLFF models through the introduction of continuous filtering convolutions~\supercite{schutt2017schnet}, which leverage smooth cutoffs to rescale the messages.
Since our main objective is to aggregate and broadcast information between a set of master nodes and the entire underlying graph, this approach is not applicable at the master node level, as the graph boundaries are not well defined: any kind of direct distance-based encoding would inherently lead to the introduction of a limited field of view given by the pairwise distribution of the training dataset. This would result in a limited transferability of the method for systems with large node delocalization.
In RANGE, we address this issue by introducing an continuous SE(3)-invariant positional encoding, where arbitrarily large distances are continuously mapped to the $[0,1]$ interval and projected into a high-dimensional space via an expansion into Gaussian radial basis functions~\supercite{schutt2017schnet}.
To verify the stability of the method, we selected a portion of the MD22 dataset~\supercite{chmiela2023accurate}, corresponding to $\sim 70$ thousand simulation frames of docosahexaenoic acid (DHA), a fatty acid consisting of 56 atoms, and trained the RANGE model on top of SchNet with a cutoff of 5\,\AA.
\begin{figure}[h!]
    \centering
    \includegraphics[width=0.9\textwidth]{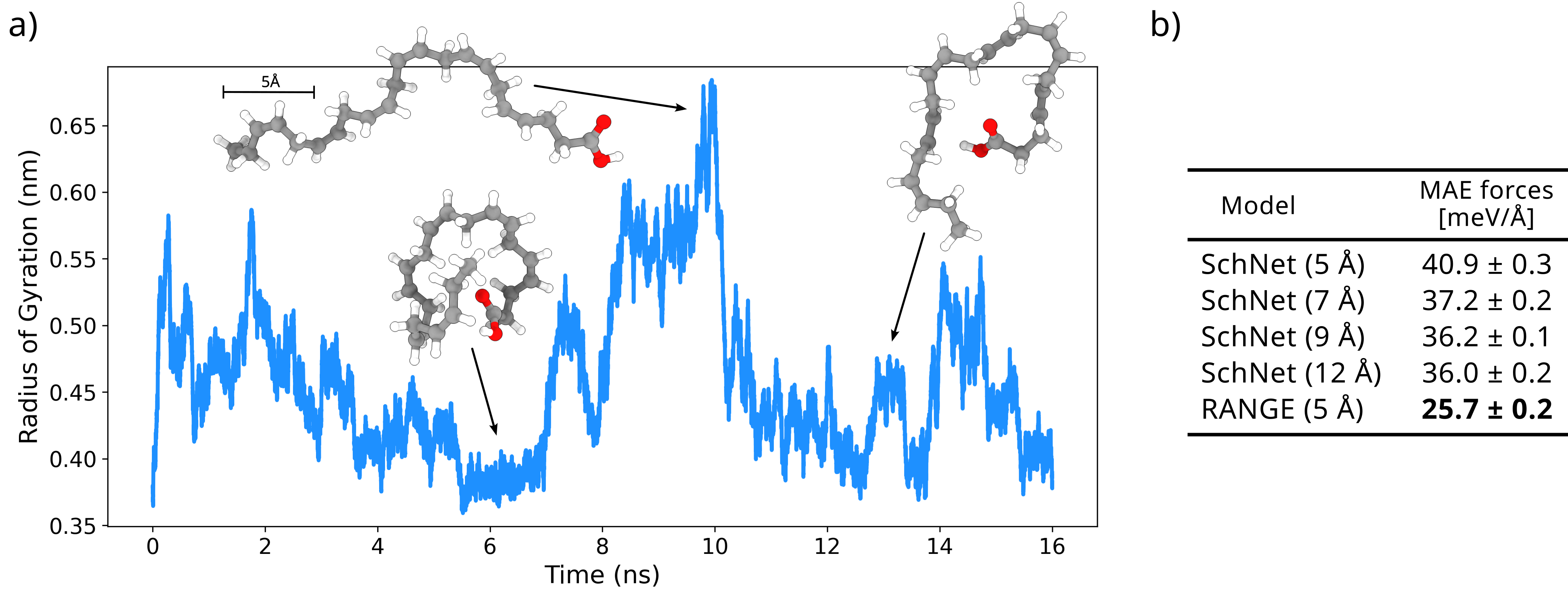}
    \caption{\textbf{Radius of gyration of DHA as a function of simulation time.} a) The radius of gyration is calculated along 16 ns of MD trajectory simulated with the RANGE architecture applied on SchNet with a 5\,\AA\, cutoff. The simulation explores different molecular conformations, realizing a full transition from a compact to an extended state and back. Representative structures from different metastable regions are reported. b) MAE forces of the SchNet baseline with different cutoff values and the RANGE model used in the simulation.}
    \label{fig:dha_traj}
\end{figure}
We report the radius of gyration during a 16 ns long MD trajectory of DHA in gas-phase, performed with the trained RANGE model, and the results of the training procedure (Fig.~\ref{fig:dha_traj}, a and b). We performed 20 independent simulations that are shown in Supplementary Fig.~\ref{sifig:dha_traj}. The regularized RANGE architecture outperforms all baseline SchNet models, despite being trained with a message-passing cutoff of only 5\,\AA. Our architecture consistently produces stable trajectories that are able to visit the complex landscape of DHA, showing complete transitions between compact and unfolded states.

\subsection*{Interpretability of RANGE}
\begin{figure}[h!]
    \centering
    \includegraphics[width=0.9\textwidth]{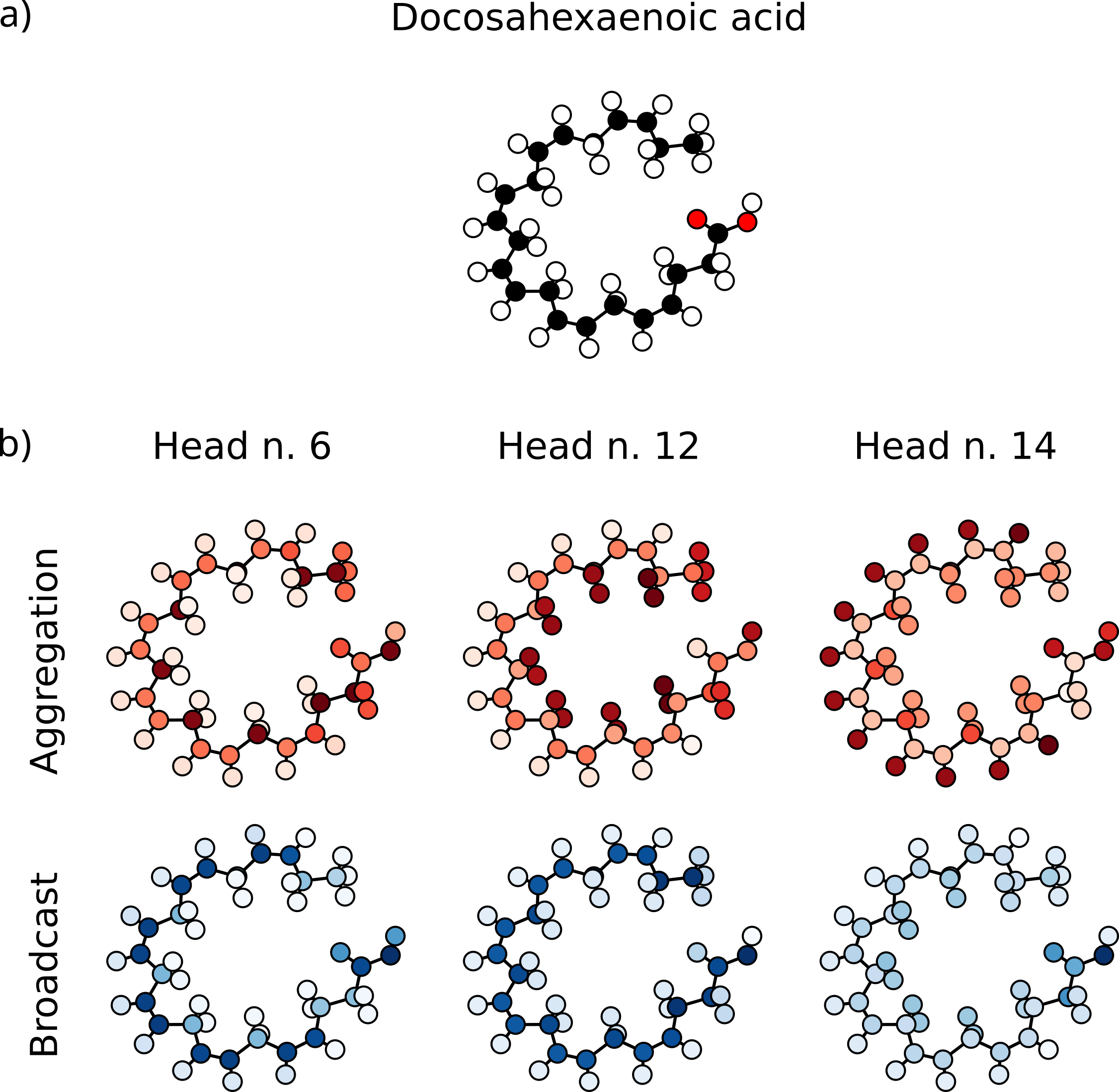}
    \caption{\textbf{Principal component of attention weights.} In a), the colors represent the atomic species (white: H, black: C, red: O). In b), the principal component of the SVD on the attention weight distribution during aggregation and broadcast for a selection of 3 attention heads is reported. Darker colors correspond to higher values.}
    \label{fig:svd_docosahexaenoic_acid}
\end{figure}
The magnitude of the self-attention weights is often used to interpret deep learning models, and understand which features are most relevant for the model outputs~\supercite{xu2015show, choi2016retain, thorne2019generating, rende2024mapping}.
The additive attention mechanism used in RANGE (Supplementary Note~\ref{sinote:range}) can provide increased flexibility with respect to the more popularized dot-product attention~\supercite{brody2021attentive}; additionally, it has been suggested that this form of attention also leads to more interpretable neural networks~\supercite{wen2022revisiting}.
Since our model preserves independence of the attention heads during each aggregation-broadcast cycle (refer to Methods section and Supplementary Note~\ref{sinote:range}), we can explore the relative importance of individual atoms within a given step. As shown in Supplementary Fig.~\ref{sifig:svd_docosahexaenoic_acid}, performing a singular value decomposition (SVD) analysis on the attention weight distribution in the DHA model discussed above reveals that each attention head typically exhibits a distinct, dominant degree of freedom, or clustering strategy.
Fig.~\ref{fig:svd_docosahexaenoic_acid} visualizes the principal component from the SVD analysis during the aggregation and the broadcast phase, mapped on the graph nodes. We report three hand-selected attention heads to illustrate the flow of information; a similar analysis for all the remaining communication channels is reported in Supplementary Fig.~\ref{sifig:pc_docosahexaenoic_acid}. As the information clustered during the aggregation phase is redistributed to the original graph nodes in the broadcasting phase, all the heads show the non-local nature of the clustering procedure and the inherently $N$-body nature of the node-node communication via virtual embeddings. This is akin to a mean-field effect, where the aggregation step outputs a weighted average of the components, and the nodes feel an effective interaction via the master node during broadcast. In this setting, each attention head produces a different learnable aggregated representation, that does not rely on predefined heuristics as typically required by clustering strategies, and allows for context-dependent weighting of information.

\section*{Discussion}
In this work, we propose RANGE: an architectural extension that can be combined with any GNN to recover long-range N-body interactions among nodes. This is achieved in a two-stage fashion via global aggregation of the information into virtual embeddings and broadcasting of the coarse-grained representations onto the nodes of the original graph.
With respect to other approaches that employ virtual aggregations, we make use of multiple, dynamically activated virtual nodes to extend the capacity of the embeddings and scale up to larger systems, and a self-attention mechanism, that has been shown to reduce oversquashing in GNNs.
We have demonstrated our framework by combining it with two popular GNN architectures, namely SchNet and PaiNN. The reported tests on accuracy and efficiency show that RANGE outperforms the baseline models in terms of accuracy and, with its linear time complexity, it outcompetes other popular solutions for the inclusion of nonlocal effects, such as Ewald-based networks, in terms of scaling.
The edge feature in our proposed model are designed in a way that guarantees the transferability across different sizes and preserves the continuity of the energy with respect to the atomic positions, a required feature in a MLFF. We report simulation trajectories for DHA that remain stable for over 15 ns, and during which the model is able to reconstruct the stable conformational states visited by this large lipid in gas-phase.
An SVD analysis of the attention weights of the virtual embeddings during the aggregation and broadcast phases reveals the presence of a single degree of freedom for each attention head, suggesting a well-defined clustering strategy; moreover, the simultaneous activation of multiple nodes spanning the entire system confirms that the distributed information is inherently N-body, leading the graph nodes to produce an adaptive mean-field effect, clustering different parts of the system during the two-phase process.

In this work, we have shown that oversquashing greatly affects the reach of MPNNs, inducing saturation in the MAE for large cutoff values. Equivariant architectures still suffer by this phenomenon, suggesting that the gains offered by including equivariant information are inherently short-range.
The results presented demonstrate the potential of attention-based virtual aggregations to improve the overall description via MPNNs of delocalized, many-body molecular systems, by creating long-range communication channels. In particular, RANGE-like implementations, that dynamically expand the capacity of the virtual embeddings via a learned regularization parameter, are able to efficiently scale up the accuracy gains to very large systems. This is achieved with a small computational overhead, constant with respect to the cutoff, and a linear scaling with system size. Future work will focus on investigating the applicability of RANGE to complex environments, such as periodic systems and solvated biomolecules, where long-range interactions play a crucial role.

\section*{Methods}
\subsection*{The RANGE architecture}
Consider a graph $\mathcal{G}$, defined by a set of $N$ nodes $\mathcal{V}$ and a set of edges $\mathcal{E}=\{\mathbf{e}_{ij}\}_{i,j=1}^{N}$, with $\mathbf{e}_{ij} \in \mathbb{R}^f$. In a standard MPNN, a learnable feature or embedding $\mathbf{h}_i^{(0)} \in \mathbb{R}^h$ is defined for every node, and sequentially updated at each interaction layer $t$ via
\begin{equation}
    \mathbf{h}_i^{(t+1)} = \upsilon_t(\mathbf{h}_i^{(t)}, \mathbf{m}_i^{(t)}),
\end{equation}
where $\upsilon_t$ is a differentiable update function; $\mathbf{m}_i^{(t)}$ is the aggregation of messages to the $i$-th node from its neighbors, defined as 
\begin{equation}
    \mathbf{m}_i^{(t)} = \bigoplus_{j\in\mathcal{N}(i)}\mu_t(\mathbf{h}_i^{(t)}, \mathbf{h}_j^{(t)}, \mathbf{e}_{ij}),
\end{equation}
where $\mu_t$ is a differentiable function and $\bigoplus_{j\in\mathcal{N}(i)}$ is a pooling operation over the neighbors $\mathcal{N}(i)$ of node $i$ designed to respect the graph symmetries. After $T$ interaction layers, a learnable readout function $\mathcal{R}(\{\mathbf{h}_i^{(t)}\}_{t=0}^T)$ is used to make predictions on the target values.
We define a master node $M$ of $\mathcal{G}$ as a virtual node that is connected with all elements in $\mathcal{V}$ via the set of edges $\mathcal{E}(M)=\{\mathbf{E}_{i}\,|\,\mathbf{E}_{i} \in \mathbb{R}^f\}_{i=1}^{N}$, with the purpose of taking long-range interactions into account by aggregating all the nodes in the graph and redistributing information. To allow for a consistent definition of the edges connecting all the graph nodes to a master node, both reside within the same space; for our application on metric graphs such as those used in MLFFs, we position each master node at the geometric center of the graph. Message-passing through $M$ consists of an aggregation and broadcast phase, as illustrated in Fig.~\ref{fig:architecture}. The former aims at harvesting information from each node embedding, collecting it in a compressed space via a GATv2-inspired multi-head self-attention mechanism~\supercite{velivckovic2017graph, brody2021attentive, wang2021egat}; the latter redistributes the coarse-grained information to each node of the graph via a self-attention mechanism that parses all the aggregated representations.
Together, aggregation and broadcast enable dynamical long-range communication between nodes. Further details on the architecture are provided in Supplementary Note~\ref{sinote:range}.

\subsection*{Data selection and preparation}
The datasets used in this work are publicly available and calculated at the DFT level of theory with PBE and PBE0 exchange-correlation functional, and corrected with many-body dispersion (MBD)~\supercite{tkatchenko2012accurate, ambrosetti2014long, stohr2016communication, mortazavi2018structure}. Further information on dataset preparation can be found in Supplementary Note~\ref{sinote:dataset}.

\subsection*{Training and MD simulations}
Models were trained and simulated using the \textit{mlcg} package~\supercite{charron2023navigating}. All models were trained for 200 epochs using a combined loss of energy and forces with the AdamW optimizer~\supercite{loshchilov2017decoupled}. Simulations were performed using a Langevin integrator at 300 K with 2 fs timestep. Further details are available in Supplementary Notes~\ref{sinote:model_training} and \ref{sinote:simulation}.

\section*{Acknowledgments}
We thank members of the Clementi's group at FU for insightful discussions and comments on the manuscript. 
We gratefully acknowledge funding from the Deutsche Forschungsgemeinschaft DFG (SFB/TRR 186, project A12; SFB 1114, projects B03, B08, and A04), MATH plus (projects AA1-6 and AA2-20), the National Science Foundation (PHY-2019745), the Einstein Foundation Berlin (project 0420815101), the Bundesministerium f\"ur Bildung und Forschung BMBF (project FAIME 01IS24076), and computing time provided on the supercomputer Lise at NHR@ZIB as part of the NHR infrastructure (project beb00040). The authors also gratefully acknowledge the Gauss Centre for Supercomputing e.V. (www.gauss-centre.eu) for funding this project by providing computing time on the GCS Supercomputer JUWELS at J\"ulich Supercomputing Centre (JSC) (project mlcg).
We are thankful to the HPC Service of FUB-IT, Freie Universit\"at Berlin, and the HPC Service of the Physics department, Freie Universit\"at Berlin; we are especially grateful to Jens Dreger for helping us with the computational setting.

\section*{Data availability}
The split files for dataset generation, and the configuration files for training and simulation will be available upon publication.
Any other data generated and analyzed for this study are available from the authors upon request.

\section*{Code availability}
The RANGE codebase will be made available upon publication.
Any additional codes are available from the authors upon request.

\newpage

\captionsetup[table]{name=Supplementary Table}
\captionsetup[figure]{name=Supplementary Figure}
\titleformat{\section}{\normalfont\Large\bfseries}{Supplementary Note \thesection:}{0.5em}{}
\setcounter{table}{0}
\setcounter{figure}{0}

\section{The RANGE architecture}
\label{sinote:range}
As illustrated in Fig.~\ref{fig:architecture} of the main text, the RANGE architecture combines a local message-passing with an aggregation of all the network nodes into a master node $M$, followed by a broadcasting that redistributes the collected information back into the single nodes, effectively realizing long-range message-passing. The details on the aggregation and broadcast phases are provided below.

\subsection{Aggregation}
Since a multi-head attention system is implemented, master nodes funnel information into $L$ $d$-dimensional spaces: the information stored in each subspace is concatenated into a $h$-dimensional vector so that $Ld = h$. The aggregated embedding is
\begin{equation}
    \mathbf{H}^{(t)} = \sigma \left( \mathbin\Vert_{l=1}^L \sum_i \hat{\alpha}_i^l A_V^l \mathbf{\tilde{h}}_i^{(t)} \right),
\label{eq:attention:aggr}
\end{equation}
where $\mathbf{H}^{(t)}\in\mathbb{R}^h$ is the embedding of $M$, $\sigma$ is an element-wise non-linear activation, $\mathbin\Vert_{l=1}^L$ represents the concatenation operator, $A_V^l : \mathbb{R}^{h} \rightarrow \mathbb{R}^{d}$ is a learnable matrix, and $\mathbf{\tilde{h}}_i^{(t)}$ refers to the $i$-th node embedding after a local message-passing iteration. Based on the conventional implementation of additive self-attention\supercite{velivckovic2017graph,brody2021attentive}, the weight $\hat \alpha_i^l$ of embedding $i$ and head $l$ is defined as:
\begin{equation}
    \alpha_i^l = (\mathbf{a}^l)^\top \text{LeakyReLU} (A_Q^l \mathbf{H}^{(t - 1)} + A_K^l \mathbf{\tilde{h}}_i^{(t)} + A_E^l \mathbf{E}_i)
\label{eq:attention:weights_aggr}
\end{equation}
\begin{equation}
    \hat{\alpha}_i^l = \text{Softmax} (\alpha_i^l) = \frac{\exp{\alpha_i^l}}{\sum_j\exp{\alpha_j^l}}.
\label{eq:attention:softmax}
\end{equation}
Here, $A_Q^l, A_K^l : \mathbb{R}^{h} \rightarrow \mathbb{R}^{d}$ and $A_E^l : \mathbb{R}^{f} \rightarrow \mathbb{R}^{d}$ are learnable matrices and $\mathbf{a}^l\in \mathbb{R}^d$ is a learnable vector. The query projection matrices $A_Q^l$ always act on the previous virtual node embedding $\mathbf{H}^{(t - 1)}$.
The edge features between master node and the graph nodes, denoted as a function of their respective distances $\mathbf{E}_i=\mathrm{RBF}(r_i)$, are carefully designed to extend the standard radial basis expansion and accommodate non-bounded distances without introducing a cutoff. We achieve this by scaling the distances between $M$ and the graph nodes by their maximum
\begin{equation}
    r_i = \frac{||\mathbf{x}_i - \mathbf{X}_M||}{\max_{j} ||\mathbf{x}_j - \mathbf{X}_M||} \quad \in [0, 1],
\end{equation}
where $\mathbf{x}_i$ denotes the position of node $i$ and $\mathbf{X}_M$ is the position of the master node, $\frac{1}{N}\sum_i \mathbf{x}_i$. The new distances are then transformed into edge features via radial basis expansion. This allows for complete transferability of the trained network across different system sizes.

\subsection{Broadcast}
In order to update the embeddings of the base graph with the aggregated information while retaining learned short-range interactions, we opted to include self-loops in the attention mechanism as follows:
\begin{equation}
    \mathbf{h}_i^{(t+1)} = \text{MLP} \left(\mathbin\Vert_{l=1}^L \left(\hat{\beta}_{i,\mathrm{self}}^l B_{V,\mathrm{self}}^l \mathbf{\tilde{h}}_i^{(t)} + \hat{\beta}_i^l B_V^l \mathbf{H}^{l(t)}\right)\right),
\label{eq:attention:bcast}
\end{equation}
where $B_{V,\mathrm{self}}^l: \mathbb{R}^{h} \rightarrow \mathbb{R}^{d}$ and $B_V^l: \mathbb{R}^{d} \rightarrow \mathbb{R}^{d}$; the latter operates on each $l$-th head representation $\mathbf{H}^{l(t)}$ separately, mantaining their independence. The attention weights are obtained with a  slight modification of Eq.~\eqref{eq:attention:weights_aggr}, by defining 
\begin{equation}
    \begin{split}
        &\beta_{i,\mathrm{self}}^l = (\mathbf{b}^l)^\top \text{LeakyReLU} (B_Q^l \mathbf{\tilde{h}}_i^{(t)} + B_{K,\mathrm{self}}^l \mathbf{\tilde{h}}_i^{(t)}) \\
        &\beta_i^l = (\mathbf{b}^l)^\top \text{LeakyReLU} (B_Q^l \mathbf{\tilde{h}}_i^{(t)} + B_K^l \mathbf{H}^{l(t)} + B_E^l \mathbf{E}_i);
    \end{split}
\label{eq:attention:weights_bcast}
\end{equation}
these are then normalized using Softmax, as defined in Eq.~\eqref{eq:attention:softmax}, to obtain the final attention weights $\hat{\beta}_{i,\mathrm{self}}^l$ and $\hat{\beta}_i^l$.
A Multi-layered Perceptron (MLP) mixes the contributions from different heads at the end of the broadcast phase, effectively integrating different classes of non-local interactions.
Remarkably, this method enables transfer of information across the system with a computational complexity that scales linearly with the number of nodes in the input graph. This is particularly advantageous when considering predictions on large systems, as it represents an improvement over standard FFT-based methods used for the treatment of long range interactions (e.g. Particle Mesh Ewald in the context of molecular dynamics), whose $N\log{N}$ scaling might represent a bottleneck during simulations of large molecules.
While we considered a single master node in the description above, this design limits the amount of relevant global information that can be aggregated without loss, thereby constraining the scalability of the model. In the following section, we will address this limitation by introducing multiple master nodes, adapting the model to tasks where the number of nodes varies significantly across the dataset.

\subsection{Spatial scalability}
When several master nodes $N_M$ with indices $I\in \{1\dots N_M\}$ are employed, each one is initialized with a different embedding $\mathbf{H}_I^{(0)}$, and Eqs.~\eqref{eq:attention:aggr} and \eqref{eq:attention:weights_aggr} become, respectively,
\begin{equation}
    \mathbf{H}_I^{(t)} = \sigma \left( \mathbin\Vert_{l=1}^L \sum_i \alpha_{iI}^l A_V^l \mathbf{h}_i \right)
\end{equation}
and
\begin{equation}
    \alpha_{iI}^l = (\mathbf{a}^l)^\top \text{LeakyReLU} (A_Q^l \mathbf{H}_I^{(t - 1)} + A_K^l \mathbf{h}_i + A_E^l \mathbf{E}_{iI}).
\label{eq:multiple_mn_aggregation}
\end{equation}
In this context, the edge features $\mathbf{E}_{iI}$ can be master node-dependent but, in order to maximize parameter sharing without sacrificing performances, the same edge features are allocated for all master nodes.
Similarly, the broadcast phase can be generalized to the case of multiple master nodes. Each $d$-dimensional portion of the output vector $\mathbf{h}_i^{(t+1)}$ can select from multiple global representations, and Eq.~\eqref{eq:attention:bcast} and the second of Eq.~\eqref{eq:attention:weights_bcast} become, respectively,

\begin{equation}
    \mathbf{h}_i^{(t+1)} = \text{MLP} \left(\mathbin\Vert_{l=1}^L \left(\hat{\beta}_{i,\mathrm{self}}^l B_{V,\mathrm{self}}^l \mathbf{\tilde{h}}_i^{(t)} + \sum_I \hat{\beta}_{iI}^l B_V^l \mathbf{H}_I^{l(t)}\right)\right)
\end{equation}
and
\begin{equation}
        \beta_{iI}^l = (\mathbf{b}^l)^\top \text{LeakyReLU} (B_Q^l \mathbf{\tilde{h}}_i^{(t)} + B_K^l \mathbf{H}_I^{l(t)} + B_E^l \mathbf{E}_{iI}).
\end{equation}
After normalizing, a regularization parameter
\begin{equation}
    \lambda_I \in 
\begin{cases} 
\{1\} & \mathrm{if}\, I = 1 \\ 
[0,1) & \mathrm{if}\, I > 1,
\end{cases}
\end{equation}
biased on the system size, rescales the contribution from each master node during broadcast by:
\begin{equation}
    \Lambda_I(n) = \lambda_I^{\gamma(n)}
\end{equation}
\begin{equation}
    \gamma(n) = (1+a_I)|\max[0,(1-n)] + \tanh(b_I) \min[1,n]|.
\label{eq:lambda}
\end{equation}
Here, $a_I$ and $b_I$ are positive trainable parameters, and $n = (N - N_\mathrm{min})/(N_\mathrm{max} - N_\mathrm{min})$ is the normalized number of nodes in the graph, with $N_\mathrm{min}$ and $N_\mathrm{max}$ being the minimum and maximum number of nodes present in the dataset during training, respectively.
While the scalar $\lambda_1$ is designed always to ensure at least one fully activated master node, the intensity of all the $\lambda_{I\neq1}$ is controlled by the factor $\gamma(n)$ as a function of the system size $n$.
Intuitively, $\gamma(n)$ should a) decrease with $n$, following the intuition that larger molecules need larger capacity per head, and b) always be greater than zero. Given these requirements, we opted for the parametric function in Eq.~\eqref{eq:lambda}, enforcing $\gamma(n)>1$ for small molecules and $\gamma(n)<1$ for large molecules, with the values $a_I$ and $b_I$ controlling this behavior.
Finally, the broadcast attention weights are rescaled as follows:
\begin{equation}
    \hat{\beta}^l_{iI} \leftarrow \Lambda_I(n) \hat{\beta}^l_{iI} \quad\quad \mathrm{for}\ I \in\{1\dots N_M\}.
\label{eq:regularization}
\end{equation}
Approaches as the one delineated in Eq.~\eqref{eq:regularization}, which aim at regularizing the overall usage of a given node in the trained model, are theoretically motivated~\supercite{giambagli2023student} and have been proven effective in real word scenarios~\supercite{liu2017learning}.

\subsection{Application to equivariant models}
Typically, SE(3)-equivariant MLFFs are designed considering 1) an invariant features representation, and 2) a set of high-order equivariant features; a mixing step is often implemented to exchange information between the two representations~\supercite{satorras2021n, schutt2021equivariant, batatia2022mace, fu2022forces}.
The RANGE aggregation and broadcast procedures, as defined in Eq.~\eqref{eq:attention:aggr} and Eq.~\eqref{eq:attention:bcast}, cannot be directly applied to SE(3)-equivariant features due to the presence of nonlinear transformations. In agreement to other designs\supercite{kosmala2023ewald, wang2024neural}, we transfer long-range information via the invariant features and possibly propagate it to the equivariant embeddings via the mixing step in the baseline model. While it is possible to explicitly incorporate higher-order equivariant features in the aggregation-broadcast scheme, this design choice maximizes computational efficiency and enables modularity in RANGE.

\section{Datasets}
\label{sinote:dataset}
All the models reported in the main manuscript have been trained on energies and forces of configurations extracted from the QM7-X~\supercite{hoja2021qm7}, AQM~\supercite{medrano2024dataset},  and MD22~\supercite{chmiela2023accurate} atomic datasets.
The labels are calculated at the DFT level of theory, with either the PBE or PBE0 exchange-correlation functional. All datasets include explicit treatment of van der Waals interactions, that are predominantly long-range, via many-body dispersion (MBD)~\supercite{tkatchenko2012accurate, ambrosetti2014long, stohr2016communication, mortazavi2018structure}.

\subsection{QM7-X}
The QM7-X dataset comprises 42 physicochemical properties calculated for $\sim$ 4.2 millions equilibrium and non-equilibrium structures of organic molecules with up to 23 atoms. These cover the set of elements that is the most predominant in biomolecules, that is H, C, N, O, S, Cl.
In order to better represent the effect of long-range interactions, a subset of QM7-X encompassing structures with more than 20 atoms was selected to train and validate the different models. The reduced dataset contains approximately 200\,000 different structures, with 99\% of all pairwise distances below 7\,\AA\, and an average of $3.4 \pm 1.3$\,\AA.

\subsection{AQM}
The Aquamarine dataset contains over 40 global and local physicochemical properties of $\sim$ 60\,000 low- and high-energy conformers of 1\,653 molecules with up to 92 atoms, both in gas phase and implicit water~\supercite{medrano2024dataset}.
In our tests, we only considered the gas phase version of the dataset and we further filtered out all structures with less than $30$ atoms. This selection led to $\sim$ 52\,000 structures with mean pairwise distance of $6 \pm 3$\,\AA. Approximately 65\% of all pairwise distances are below 7\,\AA, 83\% are below 9\,\AA\, and 95\% are below 12\,\AA.

\subsection{DHA}
We selected the portion of the MD22 dataset associated to the Docosahexaenoic Acid (\emph{DHA}), a lipid of biological interest composed of 56 atoms. Atomic and molecular properties are reported for $\sim$ 70\,000 structures.
The mean pairwise distance between the atoms of each molecule in the dataset is $6 \pm 3$\,\AA\, with 63\% of them below 7\,\AA, 81\% below 9\,\AA\, and 94\% below 12\,\AA.

\section{Model training}
\label{sinote:model_training}
All the models where trained using the combined force and energy loss:
\begin{equation}
    \mathcal{L} = \alpha \sum_{i=1}^N |E_i - E(\mathbf{X_i}; \mathbf{\theta})|^2 + \sum_{i=1}^N |\mathbf{F}_i + \nabla E(\mathbf{X_i}; \mathbf{\theta})|^2.
\end{equation}
Here, $N$ is the number of molecules, $E_i$ and $\mathbf{F}_i$ are the potential energy and forces acting on the $i$-th molecule. $E(\mathbf{X_i}; \mathbf{\theta})$ and $\nabla E(\mathbf{X_i}; \mathbf{\theta})$ are energy and forces predicted by the model, that depend on the network parameters $\mathbf{\theta}$. Finally, $\alpha$ is a scalar value controlling the relative numerical weight between force and energy contribution.
A term that acts specifically on the parameters that regulate the activation of multiple master nodes is introduced in the loss function as
\begin{equation}
    \mathcal{L}_{\text{reg}} = \sum_I \delta |\lambda_I + a_I + b_I|,
\end{equation}
where the scalar $\delta$ was set to 2.0 during all the trainings.
All models were trained on the QM7-X and AQM datasets for 200 epochs, while the training on the DHA dataset was extended to 500 epochs. The AdamW~\supercite{loshchilov2017decoupled} optimizer was used in all training, with initial learning rate of 0.0001 and a weight decay of 0.01. For the first 125 epochs, $\alpha$ was set to 0.01, and subsequently increased to 0.1. A linear scheduler was used with a gamma factor of 0.8 and learning rate step size of 19 for optimizing the model parameters, 6 for the regularization parameter $\lambda_I$, and 8 for the parameters $a_I$ and $b_I$.
In order to scale different parameter groups with different step sizes, we employed a custom implementation of the standard \textit{LinearLR} class in the PyTorch library~\supercite{paszke2019pytorch}. Model hyperparameters are reported in Supplementary Table~\ref{sitab:base_hyperparams} and Supplementary Table~\ref{sitab:rbf_hyperparams}.

\begin{table}[h]
\centering
\caption{\textbf{Training hyperparameters.} Neural network hyperparameters used for all baseline models and their RANGE counterparts.}
\label{sitab:base_hyperparams}
\begin{tabular}{lc} 
    \toprule
                                        & Training setup \\
    \midrule
    Hidden channels ($H$)               & 512 \\ 
    Number of Filters ($L$)             & 512 \\
    Interaction Blocks ($T$)            & 3 \\ 
    Activation                          & tanh \\
    Cutoff function                     & CosineCutoff \\
    Distance Expansion Basis            & Gaussian RBF~\supercite{schutt2017schnet} \\
    Master node RBF dimension           & 7 \\
    Output Network                      & MLP, 2 layers, [128,64] features \\
    Output Prediction                   & energy, forces \\
    Attention heads                     & 16 \\
    \bottomrule
\end{tabular}
\end{table}

\begin{table}[h]
\centering
\caption{\textbf{Radial basis expansion.} Dimension of the radial basis expansion used in all baseline models and their RANGE counterparts for different cutoff radii.}
\label{sitab:rbf_hyperparams}
\begin{tabular}{ccc} 
    \toprule
    Radius (\AA) & Number of RBF & Number of RBF \\
                 & SchNet        & PaiNN \\
    \midrule
    4.0  & 27  & -  \\
    5.0  & 33  & 20 \\
    7.0  & 47  & 28 \\
    9.0  & 60  & 36 \\
    12.0 & 80  & 48 \\
    \bottomrule
\end{tabular}
\end{table}

\subsection{Timing}
All time measurements were performed considering the mean training time averaged over 200 epoch. To ensure accurate and reliable evaluation of this metric, all time measurements were performed in a controlled environment: a compute node with 4 \emph{NVIDIA RTX A6000-ADA} GPUs isolated from the main compute cluster and a refrigerating system were reserved for this work in order to avoid slow downs due to over-warming. Temperature and power were constantly measured for every GPU during training as indicators of the experiments' stability. The goodness of the experimental setting is confirmed further by the low relative errors reported in Supplementary Tables~\ref{sitab:qm7x_aqm_dha_general} and \ref{sitab:aqm_painn_schnet}.

\begin{figure}[h]
   \centering
   \includegraphics[width=0.9\textwidth]{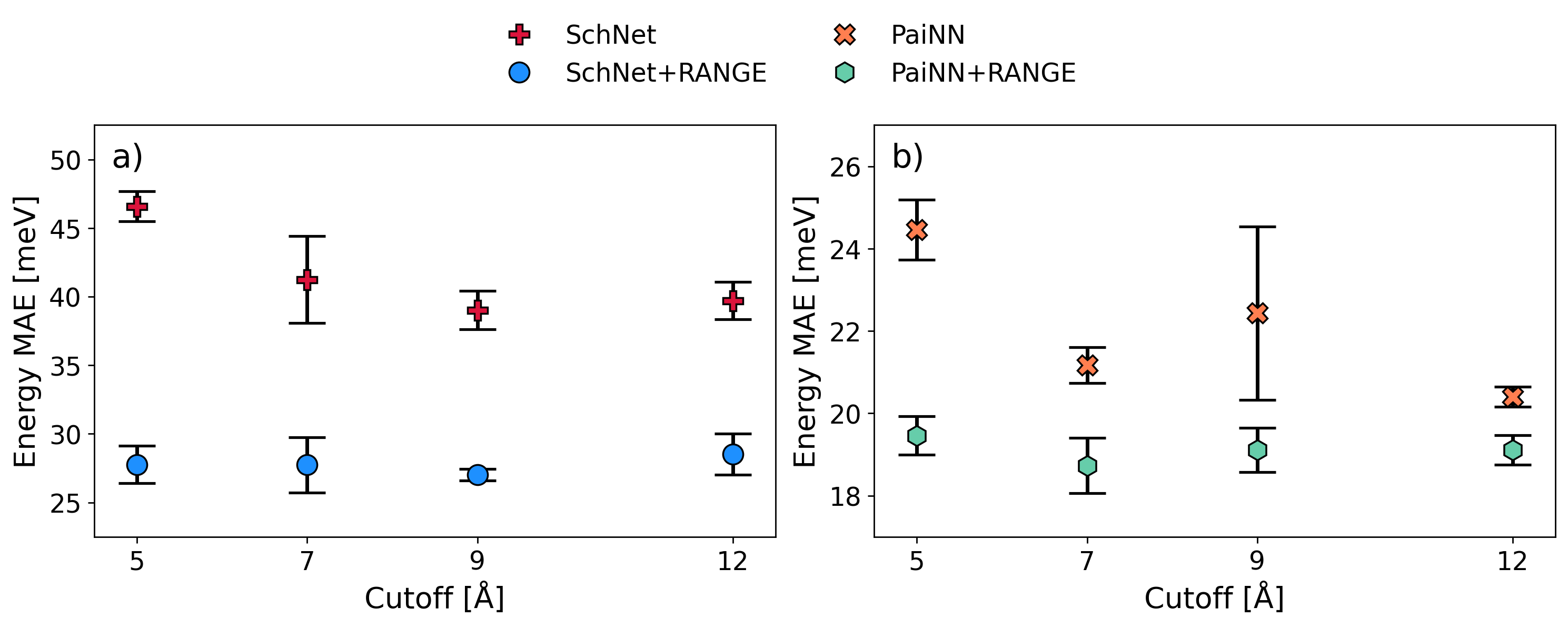}
   \caption{\textbf{Accuracy dependence on message-passing cutoff.}
    The MAE on the predicted energy of the AQM dataset is reported for a) SchNet and b) PaiNN, and the same models with the RANGE extension, as a function of the message-passing cutoff. All the reported values are averaged on 4 models independently trained with different dataset seeds.}
    \label{sifig:accuracy_timing}
\end{figure}

\begin{figure}[h]
    \centering
    \includegraphics[width=0.9\textwidth]{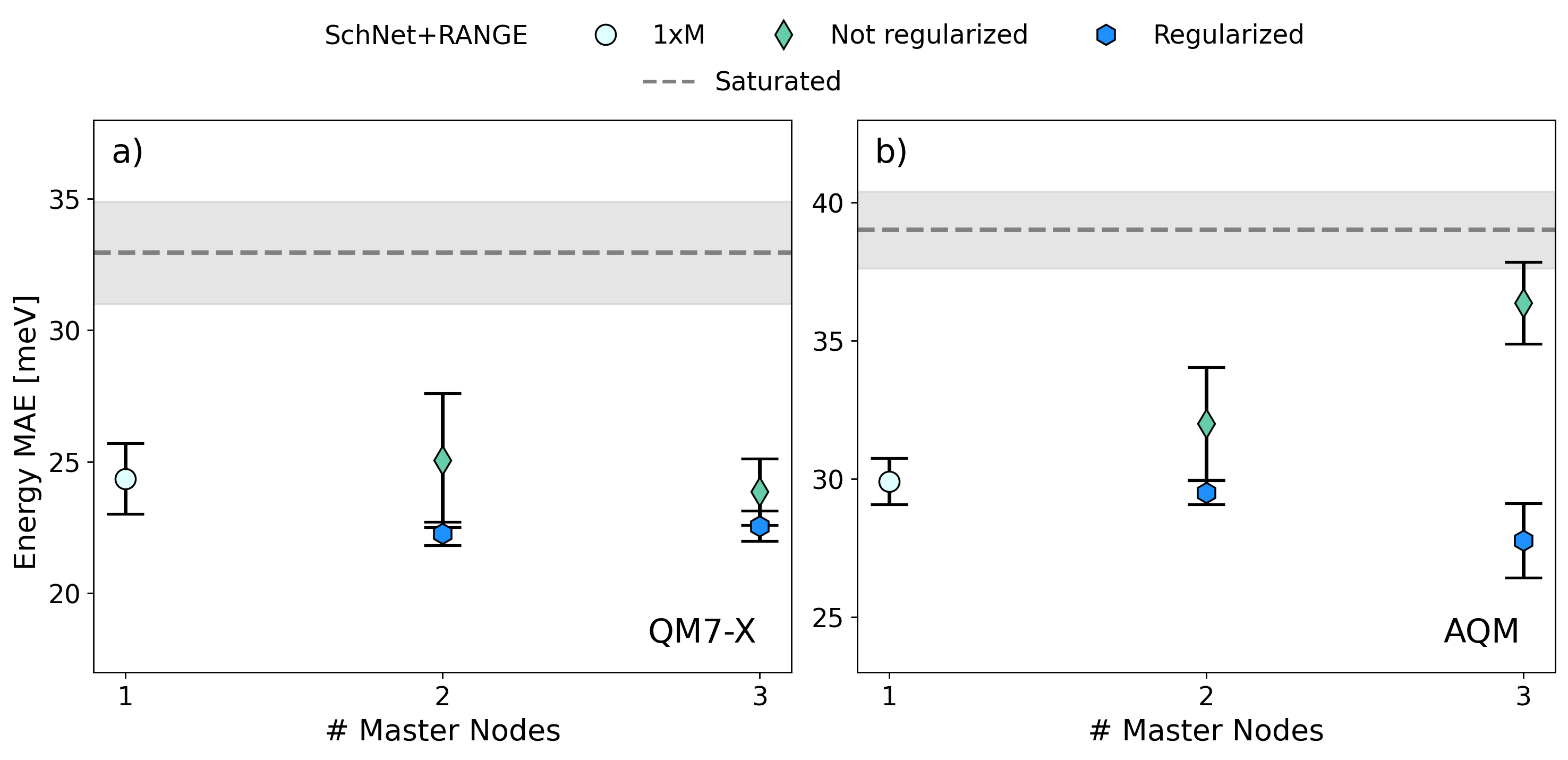}
    \caption{\textbf{MAE of the regularized and non-regularized RANGE model.} Energy MAE of the regularized and non-regularized RANGE models with different number of master nodes are reported for the a) QM7-X and b) AQM datasets. The gray line represents the lowest  MAE achieved by the baseline model upon increasing the message-passing cutoff. All the reported values are averaged on 4 models independently trained with different dataset seeds.}
    \label{sifig:not_regularized_accuracy_timing}
\end{figure}

\begin{figure}[h]
    \centering
    \includegraphics[width=0.9\textwidth]{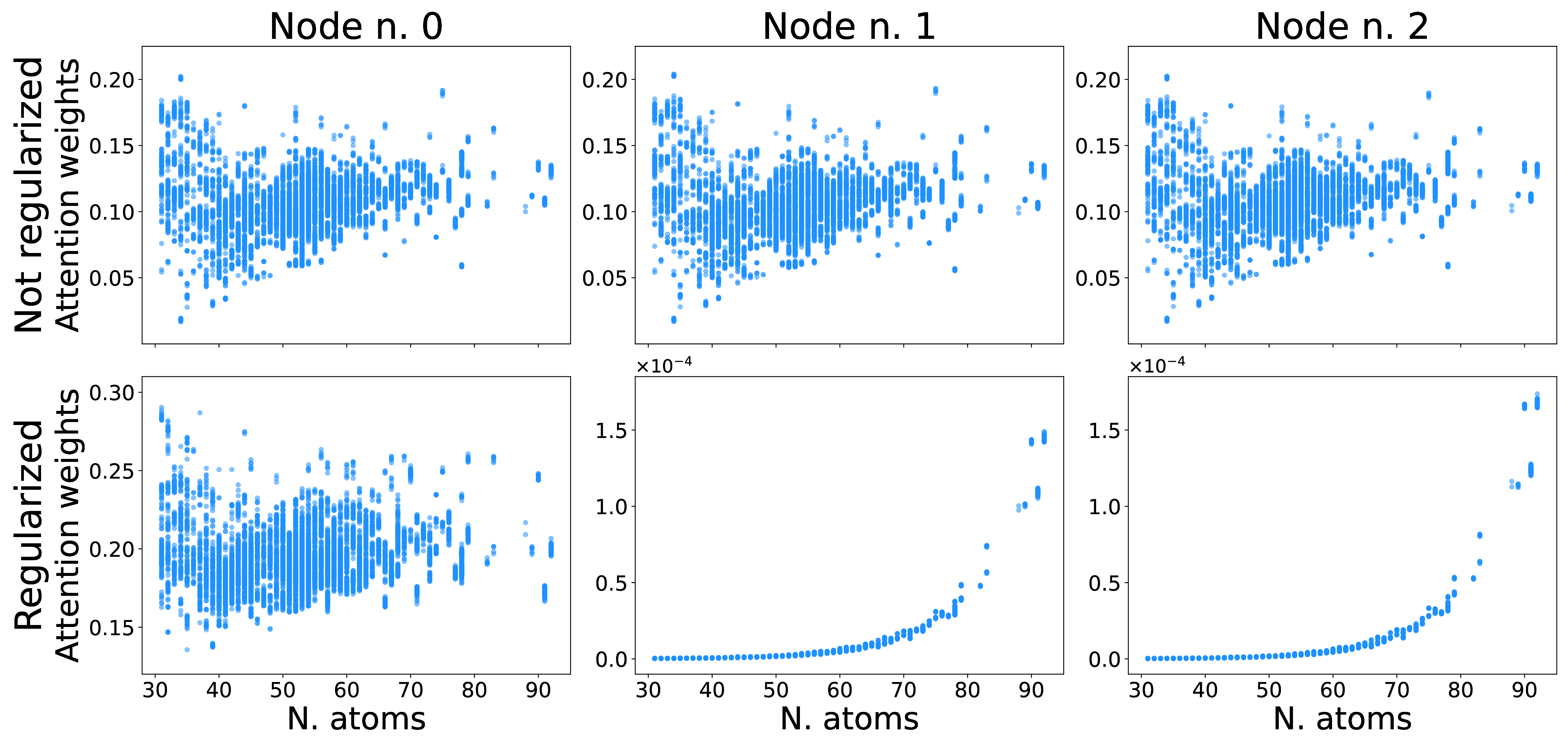}
    \caption{\textbf{Magnitude of the regularization.} Comparison of mean molecular broadcast attention weights between the non-regularized and regularized SchNet+RANGE model with 3 master nodes on the AQM dataset. The regularized model effectively reduces the relevance of nodes 1 and 2, mitigating the redundancy observed in the non-regularized model, for the smallest samples in the validation set.}
    \label{sifig:regularization_weights}
\end{figure}

\begin{table}[ht]
    \centering
        \caption{\textbf{Accuracy and training time on QM7-X, AQM, and DHA datasets.} Accuracy and training time are reported for different SchNet models, and the RANGE model with varying number of master nodes $M$ (1, 2, and 3). Non regularized RANGE models are indicated as RANGE-NR. All the reported values are averaged on 4 models independently trained with different dataset seeds. The best results are in bold lettering.}
    \label{sitab:qm7x_aqm_dha_general}
    \begin{tabular}{l*{3}{c}}
        \toprule
         Model & MAE energy & MAE forces & Training time\\
               & [meV] & [meV/\AA] & [min/epoch]\\
        \midrule
         \multicolumn{4}{c}{\textbf{QM7-X}} \\
        \midrule
         Baseline 4\,\AA & $39.2 \pm 1.8$ & $51.4 \pm 0.1$ & $0.809 \pm 0.002$ \\
         Baseline 5\,\AA & $34.6 \pm 1.3$ & $47.3 \pm 0.1$ & $0.993 \pm 0.003$ \\
         Baseline 7\,\AA & $33 \pm 2    $ & $45.0 \pm 0.2$ & $1.123 \pm 0.002$ \\
         Baseline 9\,\AA & $30.5 \pm 0.8$ & $43.9 \pm 0.2$ & $1.131 \pm 0.002$ \\
         RANGE 4\,\AA\,(1xM) & $24.4 \pm 1.4$ & $33.3 \pm 0.4 $ & $1.243 \pm 0.005$ \\
         RANGE 4\,\AA\,(2xM) & $\mathbf{22.3} \pm 0.4$ & $32.73 \pm 0.12$ & $1.316 \pm 0.005$ \\
         RANGE 4\,\AA\,(3xM)  & $22.6 \pm 0.6$ & $\mathbf{32.57} \pm 0.14$ & $1.391 \pm 0.004$ \\
         RANGE-NR 4\,\AA\,(2xM)  & $25   \pm 3  $ & $33.4 \pm 0.3$ & $1.32 \pm 0.01$ \\
         RANGE-NR 4\,\AA\,(3xM)  & $23.9 \pm 1.3$ & $33.6 \pm 0.4$ & $1.40 \pm 0.02$ \\
        \midrule
         \multicolumn{4}{c}{\textbf{AQM}} \\
        \midrule
         Baseline 5\,\AA & $46.6 \pm 1.1$ & $20.3 \pm 0.2$ & $0.831 \pm 0.002$ \\
         Baseline 7\,\AA & $41   \pm 3$   & $18.6 \pm 0.2$ & $1.257 \pm 0.002$ \\
         Baseline 9\,\AA & $39.0 \pm 1.4$ & $18.6 \pm 0.3$ & $1.550 \pm 0.002$ \\
         Baseline 12\,\AA & $39.7 \pm 1.4$ & $18.7 \pm 0.3$ & $1.791 \pm 0.003$ \\
         RANGE 5\,\AA\,(1xM)  & $29.9 \pm 0.8$ & $13.6 \pm 0.3$ & $1.212 \pm 0.005$ \\
         RANGE 5\,\AA\,(2xM)  & $29.5 \pm 0.4$ & $13.4 \pm 0.4$ & $1.250 \pm 0.006$ \\
         RANGE 5\,\AA\,(3xM)  & $\mathbf{27.8} \pm 1.4$ & $\mathbf{12.9} \pm 0.4$ & $1.284 \pm 0.006$ \\
         RANGE-NR 5\,\AA\,(2xM)  & $32   \pm 2  $ & $14.4 \pm 0.7$ & $1.241 \pm 0.002$ \\
         RANGE-NR 5\,\AA\,(3xM)  & $36.4 \pm 1.5$ & $15.1 \pm 0.3$ & $1.267 \pm 0.005$ \\
        \midrule
         \multicolumn{4}{c}{\textbf{DHA}} \\
        \midrule
         Baseline 5\,\AA & $34.9 \pm 0.3$ & $40.9 \pm 0.3$ & - \\
         Baseline 7\,\AA & $28.2 \pm 0.3$ & $37.2 \pm 0.2$ & - \\
         Baseline 9\,\AA & $25.1 \pm 0.1$ & $36.2 \pm 0.1$ & - \\
         Baseline 12\,\AA & $23.1 \pm 0.4$ & $36.0 \pm 0.2$ & - \\
         RANGE 5\,\AA\,(1xM) & $16.6 \pm 0.3$ & $26.6 \pm 0.1$ & - \\
         RANGE 5\,\AA\,(2xM) & $16.00 \pm 0.08$ & $26.0 \pm 0.1$ & - \\
         RANGE 5\,\AA\,(3xM) & $\mathbf{15.7} \pm 0.4$ & $\mathbf{25.7} \pm 0.2$ & - \\
        \bottomrule
    \end{tabular}
\end{table}

\begin{table}[ht]
    \centering
    \caption{\textbf{Accuracy and training time of SchNet+RANGE and PaiNN+RANGE on the AQM dataset.} Accuracy and training time are reported for different SchNet and PaiNN models, and their RANGE-corrected variants. All the reported values are averaged on 4 models independently trained with different dataset seeds. The best results are in bold lettering.}
    \label{sitab:aqm_painn_schnet}
    \begin{tabular}{ll*{3}{c}}
        \toprule
         & Model & MAE energy & MAE forces & Training time\\
         &       & [meV] & [meV/\AA] & [min/epoch]\\
        \midrule
         \parbox[t]{2mm}{\multirow{8}{*}{\rotatebox[origin=c]{90}{SchNet}}} & Baseline 5\,\AA & $46.6 \pm 1.1$ & $20.3 \pm 0.2$ & $0.831 \pm 0.002$ \\
         & Baseline 7\,\AA & $41   \pm 3$   & $18.6 \pm 0.2$ & $1.257 \pm 0.002$ \\
         & Baseline 9\,\AA & $39.0 \pm 1.4$ & $18.6 \pm 0.3$ & $1.550 \pm 0.002$ \\
         & Baseline 12\,\AA & $39.7 \pm 1.4$ & $18.7 \pm 0.3$ & $1.791 \pm 0.003$ \\
         & RANGE 5\,\AA & $27.8 \pm 1.4$ & $12.9 \pm 0.4$ & $1.284 \pm 0.006$ \\
         & RANGE 7\,\AA & $28   \pm 2  $ & $\mathbf{12.7} \pm 0.3$ & $1.692 \pm 0.017$ \\
         & RANGE 9\,\AA & $\mathbf{27.0} \pm 0.4$ & $12.9 \pm 0.3$ & $1.971 \pm 0.011$ \\
         & RANGE 12\,\AA & $28.5 \pm 1.5$ & $13.5 \pm 0.3$ & $1.971 \pm 0.011$ \\
        \midrule
         \parbox[t]{2mm}{\multirow{8}{*}{\rotatebox[origin=c]{90}{PaiNN}}} & Baseline 5\,\AA & $24.5 \pm 0.7$ & $8.92 \pm 0.14$ & $3.103 \pm 0.005$ \\
         & Baseline 7\,\AA & $21.2 \pm 0.4$ & $8.59 \pm 0.14$ & $4.705 \pm 0.006$ \\
         & Baseline 9\,\AA & $22   \pm 2  $ & $8.7  \pm 0.3 $ & $5.6 \pm 0.4$ \\
         & Baseline 12\,\AA & $20.4 \pm 0.2$ & $8.62 \pm 0.12$ & $6.692 \pm 0.003$ \\
         & RANGE 5\,\AA & $19.5 \pm 0.5$ & $7.68 \pm 0.17$ & $3.71 \pm 0.01$ \\
         & RANGE 7\,\AA & $\mathbf{18.7} \pm 0.7$ & $7.30 \pm 0.06$ & $5.28 \pm 0.01$ \\
         & RANGE 9\,\AA & $19.1 \pm 0.5$ & $\mathbf{7.26} \pm 0.18$ & $6.422 \pm 0.008$ \\ 
         & RANGE 12\,\AA & $19.1 \pm 0.4$ & $7.47 \pm 0.17$ & $7.24 \pm 0.02$ \\
        \bottomrule
    \end{tabular}
\end{table}

\section{Simulation details}
\label{sinote:simulation}
\begin{sidewaysfigure}[h]
    \centering
    \includegraphics[width=0.9\textwidth]{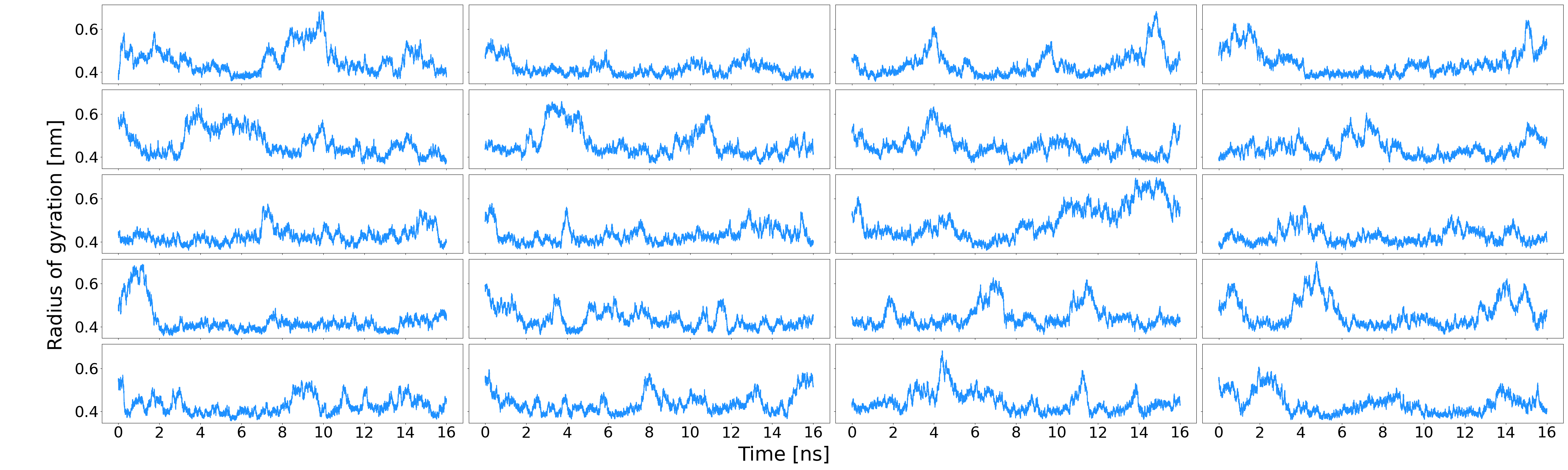}
    \caption{\textbf{Radius of gyration of DHA as a function of simulation time.} The radius of gyration is calculated along 16 ns of MD trajectory simulated with the RANGE architecture applied on SchNet with a 5\,\AA\, cutoff, across 20 independent trajectories.}
    \label{sifig:dha_traj}
\end{sidewaysfigure}

All-atom simulations of DHA were conducted using a SchNet+RANGE model with a baseline cutoff of 5.0\,\AA, 3 master nodes, and 16 attention heads for stability analysis. Each simulation was run for 16\,ns using a Langevin integrator at 300\,K, with a timestep of 2\,fs. To gather robust statistics on the conformational space exploration by each model, 20 parallel simulations were performed. Supplementary Fig.~\ref{sifig:dha_traj} presents the time series of the radius of gyration during the simulations. Notably, the model successfully explored a diverse range of DHA conformations, spanning compact and extended states.

\section{Interpretation and singular value decomposition analysis}
\begin{sidewaysfigure}
    \centering
    \includegraphics[width=0.9\textwidth]{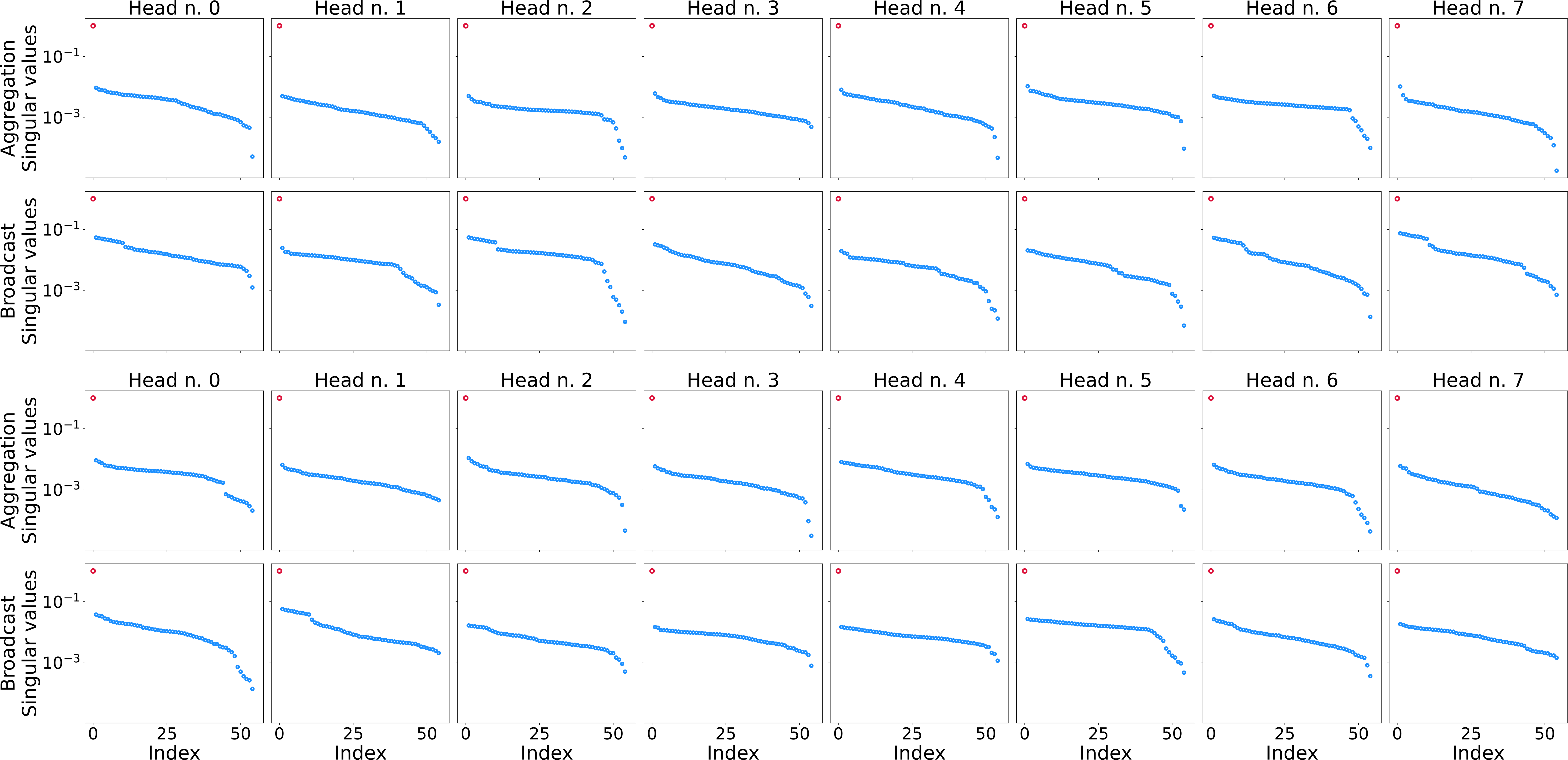}
    \caption{\textbf{Singular value decomposition (SVD) of aggregation and broadcast weights.} The SVD analysis is performed on the master node with $\lambda_1=1$. Its principal component, corresponding to the largest value, is marked in red.}
    \label{sifig:svd_docosahexaenoic_acid}
\end{sidewaysfigure}

\begin{sidewaysfigure}
    \centering
    \includegraphics[width=0.9\textwidth]{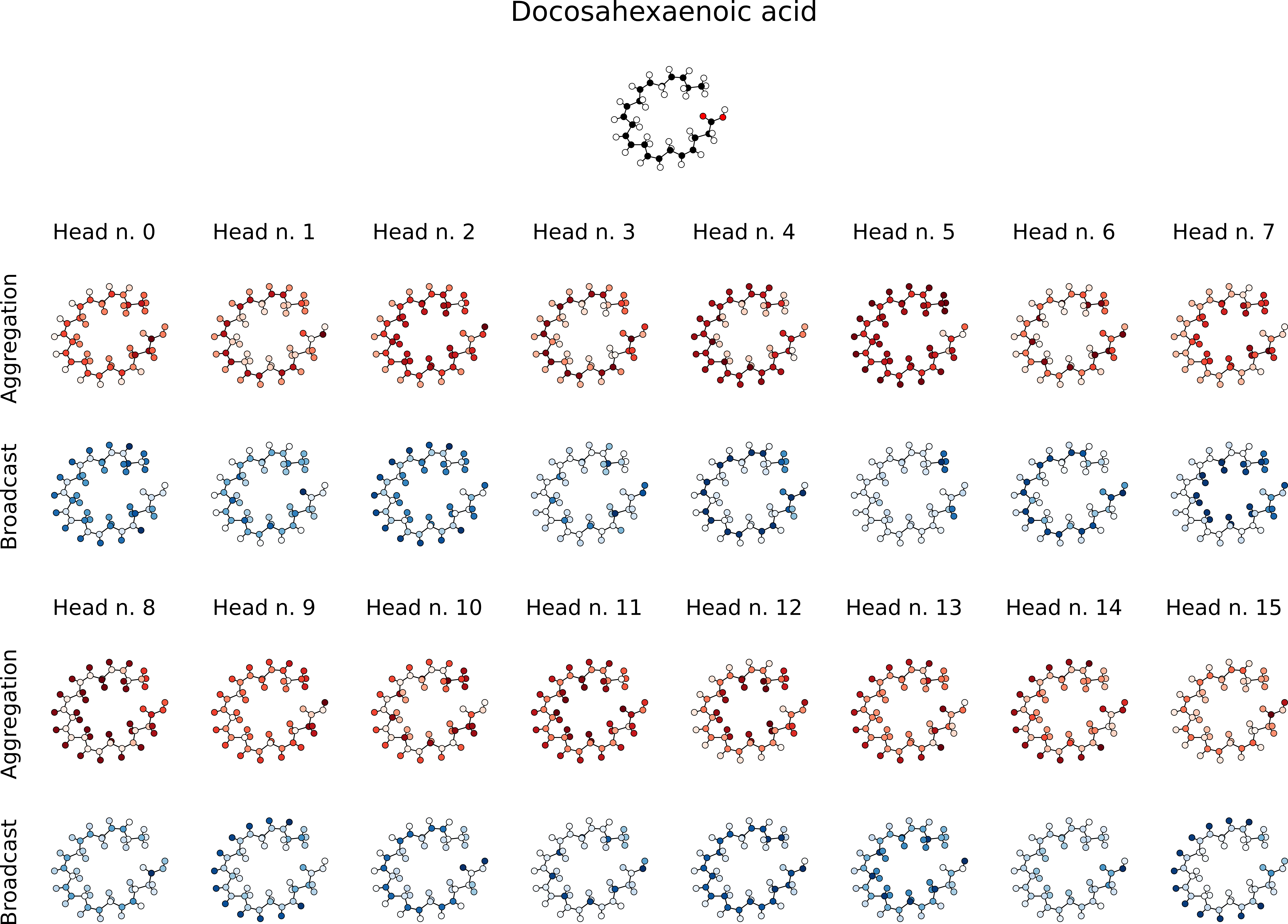}
    \caption{\textbf{Principal component of attention weights.} The colors in the top figure represent the atomic species (white: H, black: C, red: O). In the bottom figure, the principal component of the SVD on the attention weight distribution during aggregation and broadcast for all 16 attention heads is reported. Darker colors correspond to higher values.}
    \label{sifig:pc_docosahexaenoic_acid}
\end{sidewaysfigure}

For each configuration in the validation set of DHA, $V_\mathrm{DHA}$, two $N$ dimensional vector, containing aggregation and broadcast weights of the master node with $\lambda_1=1$ during the last interaction block, are stored as matrix rows to analyze the attention patterns of the RANGE model. The two matrices of size $|V_\mathrm{DHA}|\times N$ are decomposed in singular values for every attention head separately. Supplementary Fig.~\ref{sifig:svd_docosahexaenoic_acid} shows the results for aggregation and broadcast. Singular values within each matrix are normalized with respect to their maximum, highlighted in red.
A single, dominant pattern associated to an $N$-dimensional principal component emerges, and its coefficients can be mapped onto the molecular graph with a color index (Supplementary Fig.~\ref{sifig:pc_docosahexaenoic_acid}).

\clearpage

\section*{References}

\printbibliography[heading=none]

\end{document}